
\let\miguu=\footnote
\def\footnote#1#2{{\parindent=0pt\baselineskip=14pt\miguu{#1}{#2}}}

     \def\K{{\cal K}}
     \def\M{{\cal M}}
     \def\N{{\cal N}}
     \def\O{{\cal O}}
     
     \def\I{{\cal I}}
     \def\D{{\cal D}}

     \def\L{\Lambda}
     \def\g{\gamma}

\def\to{\rightarrow}
\def\real{\rm I\! R}

\def\dal{\displaystyle{{\hbox to 0pt{$\sqcup$\hss}}\sqcap}}
\def\emp{\displaystyle{{\hbox to 0pt{$0$\hss}}/}}
\def \less {\backslash}
\def \cl {\overline}

\def\tilde{\widetilde}

\def \==> {\Longrightarrow}
\def\centreline#1{\centerline{#1}}
\def\ideq{\equiv}

     \magnification=1200
     \hsize    = 165 true mm 
     \vsize    = 215 true mm  
     \parskip  = 3 true  pt plus 1 true pt minus 1 true  pt
     \baselineskip = 16 true pt plus 1 true pt minus 1 true pt

\centreline{\bf  A  Positive  Mass Theorem Based on the Focusing and
               Retardation of Null Geodesics}
\vskip 1.0 true cm
\centreline{R. Penrose}
\centreline{Mathematical Institute, University of Oxford,
24--29 St. Giles, Oxford OX1 3LB, UK}
\vskip 0.25 true cm
\centreline{R.D. Sorkin}
\centreline{Dept. of Physics, Syracuse University, Syracuse, NY
 13244--1130, USA}
\vskip 0.25 true cm
\centreline{\it and}
\vskip 0.25 true cm
\centreline{E. Woolgar}
\centreline{Jet Propulsion Laboratory, California Institute of Technology,}
\centreline{Mail Stop 301--150, 4800 Oak Grove Drive, Pasadena, CA 91109, USA}
\vskip 1.0 true cm

\centreline{\bf Abstract}
A positive mass theorem for General Relativity Theory is proved.  The
proof is 4-dimensional in nature, and relies completely on arguments
pertaining to causal structure, the basic idea being that positive
energy-density focuses null geodesics, and correspondingly retards
them, whereas a  negative total mass would advance them.
Because it is not concerned with what lies behind horizons,
this new theorem applies
in some situations not covered by previous positivity theorems.
Also, because geodesic focusing is a global condition, the proof might
allow a generalisation to semi-classical gravity, even though quantum
violations of local energy conditions can occur there.
\vskip 2 true cm
{\centreline{\it Submitted to Commun. Math. Phys.}}
\par\vfil\eject

\noindent
\centreline{\bf I: Introduction}
\vskip 0.3 true cm
\noindent
Fifteen years ago, the construction and proof of a suitable positive
mass theorem for General Relativity remained one of the major unsolved
problems of the theory.  The first proof of such a theorem was given
by Schoen and Yau,$^{(1)}$ as the culmination of a considerable effort
by several workers.  The proof they obtained was mathematically
sophisticated and  not easily accessible to many physicists.
Shortly thereafter, Witten$^{(2)}$ found a method of proof which could
be given in easily understood terms, though the rigourous mathematical
justification of the method had to await subsequent work by other
authors.$^{(3,4)}$

One might pose two questions concerning these efforts, the first
pertaining to the generality of the theorems proved.  It would seem
unreasonable to expect that a theorem could be proved even when the
spacetime is nakedly singular (after all, there is the negative mass
Schwarzschild solution) or when the non-gravitational energy of the
matter present is mostly negative (for one expects gravitational
binding energy to be negative as well, so how could the sum be
positive?). Yet one still might ask whether the assumptions
required for the existing proofs could be weakened or altered, say by
relaxing the condition of local positivity of energy or by finding a
proof which holds on spacetimes that lack complete spacelike
hypersurfaces or do not admit
Sen-Witten spinors.
It will be seen that, in fact, all these conditions can be relaxed,
although other conditions (such as the so-called generic condition, a
particularly mild assumption in the present context) have to be added.
This brings one to the second and related question, which is more
\ae sthetic in nature.  If one is interested in identifying those
features of an asymptotically flat spacetime which lead to a positive
total mass, one might like to have a theorem whose proof relies as
clearly and as deeply as possible on physically intuitive concepts.
The second question, then, is whether there exist positive mass
theorems which make more explicit use of what one might feel are the
physical mechanisms that act to ensure that the mass is positive.  The
theorem presented herein is hoped to be of this nature.

The idea is to show that negative mass spacetimes are inherently
inconsistent.  The key to the contradiction is the behaviour of null
geodesics in gravitational fields.  In 1964, Shapiro$^{(5)}$ noticed
that light rays emitted at an infinitely distant point and passing
near a positive mass should be delayed with respect to rays passing
farther from the mass, and that this effect could be used as the basis
for what is today a very accurate test of General Relativity.$^{(6)}$
For present purposes, it is the complementary observation that is of
primary interest, specifically that the rays that pass infinitely far
from a positive mass ``source'' are {\it infinitely advanced} with
respect to rays passing nearby the source.  In the negative mass case,
as one would expect, the distant rays are {\it infinitely delayed}
with respect to the nearby ones.  {}From this effect one may infer
that, in the negative mass case, there are points in arbitrarily
distant future regions of spacetime which are not causally related to
other points in arbitrarily distant past regions.  In any such
situation, there will exist null geodesics which traverse the boundary
between causally related points and unrelated ones, and in this case
it will be possible to show that one such null geodesic does so for an
infinite affine distance.  (This geodesic will pass nearby or through
the source, even though its existence is a consequence of the
behaviour of null geodesics passing arbitrarily far from it.)

Yet it is equally possible to argue that this null geodesic cannot
exist.  On one hand, it cannot have a pair of conjugate points since,
if it did, it necessarily could not remain on the
boundary between causally related and unrelated points for more than a
finite affine distance. (Recall that conjugate points are pairs of points
where some Jacobi field --- which measures the separation of neighbouring
curves of a congruence --- vanishes.$^{(7,8)}$ When this happens, {\it
focusing}
of the congruence is said to have occurred. See fig. (1).)
But on the other hand, the existence of pairs
of conjugate points along every null geodesic is a consequence of the
{\it focusing theorems}$^{(9,10,11)}$ which play such a central role in
the proofs of the singularity theorems,$^{(7,8)}$ and which are now
known to hold under very general conditions.  One thus obtains the
desired contradiction in any situation in which a focusing theorem
can be proved.

The conditions assumed by most focusing theorems express two ideas.
One is that each null geodesic should encounter curvature which
produces a converging or shearing effect on the cross-section of a
congruence of neighbouring null geodesics;
\footnote{$^*$}
{Of course, a generic curvature tensor will produce both shearing
and convergence (or divergence if the energy density is negative), but
even pure local shearing (``tidal shearing'') will produce convergence
at second order.  Thus the condition means physically that the
geodesic passes through a converging gravitational lens.}
the other is that it should not encounter enough negative
energy matter to prevent the ensuing appearance of conjugate points.
It is also necessary that the null geodesic of interest should
traverse a singularity-free region of the spacetime, which in
particular guarantees that it can be extended far enough to develop
conjugate points.  This is insured by requiring that the subset of
spacetime which is outside all event horizons be globally
hyperbolic. Incidentally, this form of causality condition means that the
positivity
theorem will apply in spacetimes with causality violations, provided those
violations are hidden behind black hole horizons (or, for that matter,
inside  white holes). The only other conditions that will be imposed pertain to
asymptotic flatness.  Asymptotic flatness is first of all used in
order that the notion of energy may sensibly be defined, but the correlative
existence of geometrical structures at infinity will also prove to
be an effective technical tool in the proof.

The theorem thus obtained is quite different in character from the
earlier theorems, which have a distinct $3+1$ flavour.
\footnote{$^{**}$}
{This flavour is less pronounced in the case of the Witten proof
which, although still proceeding on a spacelike hypersurface, can be
presented in a form where all the basic quantities which appear
(including the superpotential for the ADM energy) are 4-dimensional
in character$^{(4)}$.
}
With those theorems, one relies on the conservation of energy to argue
that, since every hypersurface that asymptotes to spatial infinity
will carry the same mass, the positivity proof may proceed entirely
on a single hypersurface; and,
in fact, the arguments used are most often
couched in terms of Cauchy data for a spacetime, rather than in terms
of the spacetime itself.  The present proof, in contrast,
is global in nature and is quite reminiscent of the sort of argument
used to prove the singularity theorems.  It is noteworthy that
both primary ingredients of the proof, namely these global techniques
and knowledge of the time delay/advance effect, were available long
before any of the existing positivity theorems were proved.

The fact that positive energy is derived herein from focusing may be
important in connection with the question of the semi-classical
stability of Minkowski space.  The usual argument for {\it classical}
stability (and also stability against quantum tunnelling) relies on
the positivity of mass as proved for stress-energy tensors that obey
local energy conditions or energy conditions integrated over spacelike
surfaces.  Such energy conditions in general fail for quantum matter
(or quantum gravitons) coupled to a classical metric, and so the usual
type of argument fails semi-classically.  However, it is conceivable
(and presently very much an open issue) that energy conditions
integrated along null geodesics may hold in appropriate circumstances,
even in the presence of quantum matter.  If such semi-classical
focusing theorems could be established, they would lead to an
extension of the positive energy theorem presented herein which would
be valid in the context of the semi-classical Einstein equation.

In the presentation that follows, Section ${\rm II.1}$ states
the theorem in a general form, relying on the assumption that null
geodesics focus; {\it i.e.} develop conjugate points.  Section ${\rm
II.2}$ discusses sufficient conditions under which every infinite null
geodesic will in fact focus, these being the conditions alluded to
above.  Section ${\rm III}$ discusses the behaviour of null geodesics
propagating in the asymptotic regions of spacetime. The utility for
this purpose of solutions of the Hamilton-Jacobi equation is the
subject of Section ${\rm III.1}$, while Section ${\rm III.2}$ contains
the calculations themselves.  One can show that an important property
of null geodesics from a fixed initial endpoint at past null infinity
is governed by the sign of a certain combination of the asymptotic metric
coefficients --- if this quantity is positive, the geodesics form a
sequence receding into the distant future and, if it is negative, they
form a sequence converging upon spatial infinity.  The quantity that
governs this behaviour is a component of the spacetime's 4-momentum, as
shown in the appendix.  Section IV gives the formal
proof of the positivity theorem.  Section V points out some purposes
for which the present theorem is more general than the others, and
contains some speculations
on possible further generalisations of the result, particularly concerning the
cases of null and zero 4-momentum.

Many of the ideas herein grew out of discussions following a seminar
given by one of the authors (R.P.) at Syracuse University in the
Spring of 1990.  Subsequently, two different but related approaches to
proving the result described in this paper were found and
have been reported elsewhere.$^{(12,13)}$  This work is intended to
give a full proof of the results stated in those previous works and to
generalise them.  For an early statement of some of the ideas behind
this proof see Ref. (14) and the related
ideas in Ref. (15); for an alternative development of part of the
work, see Ashtekar and Penrose.$^{(16)}$

The conventions adopted are those of the Landau-Lifshitz Spacelike
Convention used in Refs. (7, 17, 18).  Specifically, the signature
of the metric is taken to be $(-,+,+,+)$, and the Ricci tensor is
defined so that it is the positive sign of timelike and
null components of $R_{ab}$ which leads to focusing (see Subsection ${\rm
II.2}$ for a more complete discussion). Small roman letters are used
for abstract indices, while small greek indices refer to components in
chosen coordinates, usually some quasi-Cartesian system near infinity
as  in Section III. {\it Spacetime} means a manifold (Hausdorff,
paracompact,
4-dimensional except where higher-dimensional generalisations are
discussed) with a Lorentzian metric; it may possess singularities (provided
they are suitably attired; again,
see subsequent discussion). A {\it causal curve} in spacetime will
generally refer to a curve-without-parametrisation --- the image of a
continuous Lipshitz map from an interval of
$\real$ into spacetime (the Lipshitz
condition is automatic for causal curves). The {\it chronological
({\rm resp.} causal) future} $I^+({\cal X})$ {\it ({\rm resp.}
$J^+({\cal X})$)} consists of all
points joined to the point ${\cal X}$ or to points in the set ${\cal
X}$ by future timelike ({\it resp.} future causal --- future timelike or null)
curves
from ${\cal X}$. Single points are considered curves of zero length,
and therefore $p\in J^+(p)$ but $p\notin I^+(p)$.
Lastly, ${\cal A} \less {\cal B}$ denotes the complement of the
set ${\cal B}$ in
the set ${\cal A}$, {\it i.e.} those points of ${\cal A}$ that are not in
${\cal B}$.

\vskip 0.5 true cm
\noindent
\centreline{\bf II: Focusing and Positivity}
\vskip 0.3 true cm
\noindent
\line{\underbar {II.1: Statement of the Theorem:}\hfill}
\vskip 0.3 true cm
\noindent
The result to be proved refers to spacetimes which are asymptotically
flat in a sense to be made precise in Section III (see especially
equations ${\rm (III.2.1)}$ and ${\rm (III.2.2)}$ and the
discussion following equation (A.3) in the appendix).
The spacetime manifold $\M$,
with Lorentzian metric $g_{ab}$, will be regarded as embedded in a
larger manifold $\widetilde\M$ (with conformally related metric
${\widetilde g}_{ab}$) in order that one may refer to points at
infinity in the usual manner.$^{(18)}$ Thus, one defines the
boundary $\I$ of $\M$ in $\widetilde\M$, and its disjoint subsets,
$\I^+$ ($\I^-$) = {\it future (past) null infinity} and $\{i^0\}$ =
{\it spatial infinity}.
Note that $\widetilde\M$ is taken to extend slightly ``beyond
infinity'' in order that it can be a manifold --- albeit not a $C^\infty$
one --- even
at $i^0$.

Notice that $\I$ is the union of the null geodesics\footnote{$^*$}
{That is, null geodesics with respect to ${\widetilde g}_{ab}$.
Recall that conformally equivalent metrics possess the same null
geodesics. }
emanating from $i^0$.
Notice also that $\I$ represents only a single asymptotic region of
$\M$; if there are others, they will be ignored, and only the mass
associated to the one whose ``boundary at infinity''
is $\I$ will be studied.  Thus $\I^+$, for example,
consists of the ideal endpoints of out-going null geodesics which reach
arbitrarily great radii in the asymptotic region in question.
One also defines the {\it domain of outer
communications} ${\cal D}=I^+({\cal I}^-)\cap I^-({\cal I}^+)$
(the set of events which ``can communicate with the asymptotic region'';
an example is depicted in fig. (2)).
It will be assumed that ${\cal D}$ satisfies the following conditions:

\item{$\bullet$}{Every infinite null geodesic in ${\cal D}$ possesses a
pair of conjugate points.}

\noindent
(In fact, this condition can be weakened to one pertaining only to
achronal null geodesics, as will be seen later.)

\item{$\bullet$}{${\cal D}\cup {\cal I}$ is {\it globally hyperbolic}
as a subset of $\widetilde\M$; a
globally hyperbolic set being one which is {\it strongly causal}
and which contains $\langle\langle p,q\rangle\rangle :=
J^+(p)\cap J^-(q)$ as a compact subset,
for each $p$ and $q$ in the set.}

\noindent
The conditions for asymptotic flatness will be taken to hold on the
intersection $\N$ of some neighbourhood of $\I$ with ${\cal D}$.  One
has then:

{\narrower\smallskip\noindent
{\underbar {Theorem II.1.1:}}
The 4-momentum of ${\cal M}$ is future-causal.\smallskip}

Equivalently, the theorem states that the ADM energy is non-negative in
every rest-frame.  Notice that the theorem as presented above does not
explicitly assume Einstein's
equations. Rather, those features of the curvature which are essential
for the theorem to hold have been extracted and posed as the
assumptions.  Positivity of mass emerges as a property of general
Lorentzian manifolds that focus null geodesics and have an adequate
asymptotic flatness property.  Upon assumption of Einstein's equations,
one can replace the conditions on the curvature by conditions on the
matter tensor.  In the next subsection, it will be seen that
reasonable conditions on the matter tensor act to enforce the
conjugate point condition.
It may also be appropriate to mention here that such causality
conditions as will be needed will apply only to $\cal D$; there is no
need to forbid, for example, closed timelike curves hidden behind
event horizons.

\par\vfill\eject
\line{\underbar {II.2: Focusing Theorems and the Existence of Conjugate
Points:}\hfill}
\vskip 0.3 true cm
\noindent
The condition requiring the existence of conjugate points may be
imposed by assuming instead other conditions which suffice to imply
the necessary focusing, as mentioned in the introduction.  The
traditional way to do this has been to require that the energy density be
pointwise non-negative, and to assume in addition that the curvature
tensor satisfies
the so-called generic condition defined below.
For details, the reader may consult Refs. (7,8,19).
However, recent work shows that geodesic focusing occurs under
conditions considerably more general than originally known.  An example
of such conditions would be that
along every infinite null geodesic the integral condition sometimes (somewhat
inappropriately) called the Averaged Weak Energy Condition hold, with
every such geodesic encountering a point of generic curvature.  Similar
conditions have been used in recent discussions of causality violating
spacetimes$^{(20)}$ and are drawn essentially from work of
Tipler$^{(9)}$. However, even Tipler's conditions can be weakened, as
was shown by Borde$^{(10)}$, who proves the following theorem.

{\narrower\smallskip\noindent
{\underbar {Borde's Focusing Theorem 2:}}
Let $\gamma$ be a complete causal ({\it i.e.} null or timelike) geodesic
with affine parameter $t$ and tangent $T^a$; and let
$T_{[a}R_{b]cd[e}T_{f]}T^cT^d\neq 0$ somewhere on
$\gamma$.  Suppose further that for each $\epsilon>0$ there exists $B>0$
such that, for every pair of affine parameter
values $t_1<t_2$, there
exists an interval $I_1$ with length $\ge B$ and endpoints $ < t_1$,
   and an interval $I_2$ with length $\ge B$ and endpoints $ > t_2$,
such that
$$
  \int\limits_{t'}^{t''}R_{ab}T^aT^b\ dt \ge -\epsilon \quad \forall
  t'\in I_1\quad ,\quad \forall t''\in I_2\quad .
$$
Then $\gamma$ contains a pair of conjugate points. \smallskip}

\noindent
Also, K\'ann\'ar$^{(11)}$ recently has produced a variant of this
theorem which imposes an integral condition directly on components of the
Riemann tensor instead of on components of the Ricci tensor.

A spacetime in which $T_{[a}R_{b]cd[e}T_{f]}T^cT^d\neq 0$ somewhere on
every geodesic (with $T^a$ tangent to the geodesic)
is said to obey the {\it generic condition}. For
present purposes, it suffices that this hold on every {\it null}
geodesic, and the word {\it generic} will be used in that sense.
Satisfaction of this condition is not at all a strong assumption,
being indeed generic in a suitably rigourous sense.$^{(19,21)}$
This last fact means in particular that, even for spacetimes which do
not satisfy the condition, there should be nearby generic
spacetimes with almost the same total mass.  Thus a proof of
positivity of mass for generic spacetimes should imply the same result
for arbitrary ones.

The motivation for studying general conditions under which focusing
might occur came from attempts to analyse the singularity theorems
(and more recently the so-called chronology protection
conjecture$^{(22)}$ forbidding formation of closed timelike curves) in
the context of classical gravity coupled to quantum fields (including
the graviton field).$^{(23)}$ The present work raises in addition the
possibility of an application to proving a positive energy theorem in
this same ``semi-classical'' context.

If integral
energy conditions of the sort used in Borde's theorem were to hold for
arbitrary states of arbitrary quantum fields in arbitrarily curved
spacetimes then the proof of positivity of mass presented herein would
generalise straightforwardly to the case of semi-classical gravity.
Although some encouraging results in this direction are
available\footnote{$^{**}$}
{In this connection, it might prove important that the proof given in
Section IV below deals with a null geodesic which is achronal by
construction.  In deriving the required contradiction it therefore
suffices to know that every {\it achronal} null geodesic has a pair of
conjugate points (and therefore is self-contradictory).  Hence it
would suffice to have available an integral energy condition known to
hold only for achronal null geodesics, as in Ref. (24).  Notice in
particular that although, for compact spacetimes, the Casimir energy
may be negative, this need not imply that the integrated energy
conditions fail on achronal null geodesics.}
(in particular the theorems of Refs. (24, 25) and the indications
from recent work of Ford and of Roman that the uncertainty principle
places strong constraints on the negative energy distributions arising
from quantum fields in flat and certain curved spacetimes$^{(26)}$),
it now appears that integral energy conditions can be violated by
{\it test} quantum fields in certain curved spacetimes.$^{(24)}$
However, no such violation
is known in the presence of the
semi-classical Einstein equation, and even in situations where one
would expect that approximate equation to fail (such as for an exploding
black hole) the obvious candidates for ``non-focusing'' null
geodesics turn out, on closer consideration, to be likely to possess
pairs of conjugate points.

Finally, it is noteworthy that focusing theorems can be generalised
to certain metrics that fail to be $C^2$, for example ones with
shock wave discontinuities in the curvature or the connection.$^{(27)}$
This makes it likely that a version of
Theorem ${\rm II.1.1}$ will still hold in such cases, thereby
providing a positivity result appropriate to such generalised
solutions of Einstein's equation.
\par\vfill\eject
\noindent
\centreline{\bf III: Null Geodesics in a Neighbourhood of Infinity}
\vskip 0.3 true cm

\noindent
{\underbar {III.1: The Phase Delay and the Retarded and Advanced
           Time Functions}}
\vskip 0.3 true cm

\noindent
In 1960, Pleba\'nski$^{(28)}$ published a study of the behaviour of null
geodesics propagating in ${\cal O}(1/r)$ weak gravitational fields.
He used his results to compute light bending, but did not explicitly note the
time delay effect,
nor did he comment on the complementary effect,
{\it i.e.} the infinite time advance of null geodesics as
one approaches infinity in a positive mass spacetime. His method of
solving the Hamilton-Jacobi equation  perturbatively
about a flat background metric (namely that entailed by asymptotic flatness)
will be employed here.
The analysis of this section will be valid only on a suitable
neighbourhood of infinity, one on which an appropriate background
structure may be defined, as detailed in the next subsection.

Fix a generator $\L^-$ of $\I^-$, let $p$ be some point on $\L^-$ ,
and consider the set of all null geodesics originating at $p$.
This set is a subset of the non-rotating congruence of all null
geodesics whose initial endpoints lie on $\Lambda^-$. Since this is a
non-rotating congruence, it can be described by a single scalar function
$S$, which is known by many names, the Hamilton-Jacobi function, the
eikonal, the phase (of the waves carried along by
the congruence), {\it etc}.  It obeys
$$
g^{\mu\nu}\partial_{\mu}S\partial_{\nu}S=0\quad .\eqno{\rm (III.1.1)}
$$
Its significance is that the null geodesics leaving $\Lambda^-$ generate
the level surfaces of $S$, so knowledge of $S$ defines in a succinct manner
the set of null geodesics from each initial point $p\in\Lambda^-$.
When $S$ is normalised so that it reduces on $\L^-$ to a suitably
defined affine parameter, it will be called the advanced time function
and denoted by $S^-$.

Consistent with asymptotic flatness, let there exist quasi-Cartesian
coordinates $\{ x^{\mu} \}$ in a neighbourhood of infinity, by which is
meant coordinates in which the (inverse) metric takes the form
$$
 g^{\mu\nu}=\eta^{\mu\nu}+h^{\mu\nu}\quad ,\quad h^{\mu\nu}={\cal O}(1/r)
 \quad ,\eqno{\rm (III.1.2)}
$$
where $t=x^0$, $r^2=x^jx^j$ (with $j$ running from $1$ to $3$)
and $\eta^{\mu\nu}$ is the (inverse) Minkowski metric, having components
${\rm diag}(-1,1,1,1)$.  The precise sense in which ${\cal O}(1/r)$
is to be understood will be described shortly.
Then one may expand $S$ about its Minkowski space form as
$$
 S=\eta_{\mu\nu}k^{\mu}x^{\nu}+\Delta S\quad ,\quad {\rm
 where}\quad \eta_{\mu\nu}k^{\mu}k^{\nu}=0\quad ,\eqno{\rm
 (III.1.3)}
$$
with $k^{\mu}$ having constant components in the $x^{\mu}$ system.
The Hamilton-Jacobi equation ${\rm (III.1.1)}$ becomes
$$
 \eqalign{
 k^{\mu}\partial_{\mu}\Delta S =& -{1\over 2} h^{\mu\nu}k_{\mu}k_{\nu}
 +{\cal E}(x^{\mu})\quad,\quad
  k_{\mu}:=\eta_{\mu\nu}k^{\nu}\quad ,\cr
 {\cal E}(x^{\mu})=&-{1\over 2}\big \{ \eta^{\mu\nu}\partial_{\mu}\Delta S
 \partial_{\nu}\Delta S +h^{\mu\nu}\partial_{\mu}\Delta S \partial_{\nu}
 \Delta S+2h^{\mu\nu}k_{\mu}\partial_{\nu}\Delta S \big \}
 \quad,\cr}\eqno{\rm (III.1.4)}
$$
where ${\cal E}$ represents an error term of second order in
$h^{\mu\nu}$ which will vanish like $1/r^{1+\epsilon}$ or faster as
$r\rightarrow\infty$.
Note the definition of $k_{\mu}$, which implies
that it also has constant components.
Neglecting the error term, one may integrate equation
${\rm (III.1.4)}$ to yield
$$
 \eqalign{\Delta S(t,x^j) =& -{1\over {2k^0}}\int\limits^t_{t_0}dt'
 \int d^3x'\ k_{\mu}k_{\nu}h^{\mu\nu}(t',{x^j}')\cr
 &\qquad\delta^{(3)} (x^j - {x^j}'-(t-t'){\hat n}^j) + C(t_0) \ ,
 \cr}
\eqno{\rm (III.1.5)}
$$
where ${\hat n}^j= k^j/k^0$ is the unit vector in the $k^j$-direction
and $C$ is a constant of integration.
When $t_0$ is taken to negative infinity with a suitable choice of
integration constants,\footnote{$^*$}
{In order that $\Delta S$ have a limit as $t_0 \to -\infty$, $C(t_0)$
generally will have to diverge logarithmically with $t_0$, {\it cf.}
equations (III.2.9) and (III.2.11) below.
Note that $S^-$ can also be
described as that solution of the Hamilton-Jacobi equation which
agrees on $\L^-$ with a certain Bondi time coordinate, as defined in
Ref. (29).
}
$S=S^-$ describes a plane wave emanating from the generator
$\Lambda^-$ of ${\cal I}^-$.  Note that the support of the integrand
in equation ${\rm (III.1.5)}$ is along the Minkowskian-null line
through $(t,x^j)$ parallel to $k^\mu$.  In what follows,
$k^\mu$ will be taken to have components (1,0,0,1), so that
$\eta_{\mu\nu} k^\mu x^\nu = z-t$ describes a plane wave propagating
in the $z$-direction.

Analogous to $S^-$ is another function, the retarded time function,
$S^+$ which will also be needed in what follows.  It furnishes a
function on spacetime which is constant along each null geodesic whose
future endpoint lies on $\Lambda^+$, the generator of $\I^+$ that
``continues'' $\Lambda^-$.  This function also obeys the
Hamilton-Jacobi equation ${\rm (III.1.1)}$ and the resulting equations
${\rm (III.1.3)}$ with a $\Delta S$ given (approximately) by ${\rm
(III.1.5)}$ with the same $k^\mu$ but the integral taken from $t$ to
$\infty$ rather than from $-\infty$ to $t$.

An important caveat is that $S^+$ (like $S^-$) is in general not
everywhere well defined.  The problem is that null geodesics focus, so
two or more null geodesics that have different endpoints on
$\Lambda^+$ may pass through some given point $x$ in spacetime,
assigning to that point two or more different values of $S^+$.  However,
these values always have a lower bound, since null geodesics from
any fixed $x$ in spacetime cannot arrive at ${\cal I}^+$ arbitrarily
early. The greatest lower bound of $S^+$ at $x$ is physically
significant in that it gives the earliest (retarded) time of arrival
at $\Lambda^+$ of any null geodesic (and hence any signal
whatsoever) from $x$.  In what follows, however, one is able to
limit consideration to null geodesics contained within the asymptotic
region, and for them the (obviously single-valued) approximation to $S$
given by (III.1.5) will suffice for present purposes\footnote{$^{**}$}
{Of course, from a fixed point of ${\cal I}^-$, unphysical (but causal)
signals may reach any generator of ${\cal I}^+$ arbitrarily early by
passing through $i^0$. Less trivially, one may
see that causal curves in spacetime may reach $\Lambda^+$ arbitrarily
early from any point of ${\cal I}^-\less \Lambda^-$ by smoothing out the
aforementioned unphysical curves so as to avoid $i^0$,
and the retarded time thus is unbounded below on approach to such
points.
See particularly the
discussion of equation ${\rm (III.2.19)}$, the remark after the proof of
Theorem ${\rm II.1.1}$ in Section IV, and fig. (3).
Communication involving physical signals
from $\Lambda^-$ to $\Lambda^+$ is key to our argument; for this case, see
Lemma ${\rm III.2.1}$.}
(in fact, all that is needed is that the error made by using (III.1.5)
be uniformly bounded on neighbourhoods of infinity of the sort
discussed in Section ${\rm (III.2)}$).

It is worthwhile to interpret the preceding equations before
proceeding. The phase of a hypothetical (massless) plane wave that
couples only to the background flat metric is
$\eta_{\mu\nu}k^{\mu}x^{\nu}$.  Equation ${\rm (III.1.5)}$ may be
thought of as giving the {\it phase delay} experienced by a wave
propagating through a refractive medium, following to zero$^{\rm th}$
order of approximation the Minkowskian-null direction $k^{\mu}$, but
experiencing refraction due to its coupling to $h^{\mu\nu}$.  To
compute null trajectories to {\it first} order of approximation, one
must note that photons propagate so as to keep the true phase $S$ of
equation ${\rm (III.1.3)}$ constant, not the background phase, so one
must consider equations ${\rm (III.1.3)}$ and ${\rm (III.1.5)}$
together.  With the choice of $k^{\mu}$ as having components
$(1,0,0,1)$, the equation $S= const.$ determines a surface of constant
phase ruled by null geodesics that would be described in 3+1-language
as normally outbound from a coordinate 2-plane $z=z_0$ in the
asymptotic region at $t=t_0$.  Then $\Delta S$ describes the amount by
which this surface differs from one ruled by lines null in the
background metric $\eta_{\mu\nu}$.
Thus, one may use $\Delta S$ to extract the {\it time delay}$^{(5,6)}$
(possibly negative) due to
$h^{\mu\nu}$, experienced by a plane wave evolving in the asymptotic region
from the initial plane surface. To see this, one may ask at what value
of coordinate time $t$ does a generator of the $S=const.$ surface reach
a point of fixed spatial coordinate $x^i$, and what would be the value
of $t$ if $h^{\mu\nu}$ were neglected. The resulting time delay is given
by $\Delta t = (1/k^0) \Delta S$, where $k^0$ is the projection of $k^{\mu}$
onto the normal to the surfaces of constant time. If this normal is $u_{\mu}=
\partial_{\mu}t$, one may write the formula
$$
\Delta t = {{\Delta S}\over {k^{\mu}u_{\mu}}} \quad ,\eqno{\rm
(III.1.6)}
$$
describing the Shapiro time-delay effect.  This is somewhat implicit,
as $\Delta S$ depends on $\Delta t$ but, consistent with the first
approximation, the final time required in the evaluation of $\Delta S$
may be taken to be the time at which the Minkowskian-null line tangent
to $k^{\mu}$ has spatial coordinates $x^i$.

Note that there is an interpretation in physical terms associated with
the use of the first order approximation described above.  It is the
approximation in which account is taken of the time delay experienced
by null geodesics, but no account is taken of their consequent
bending.  Such an approximation clearly is valid for geodesics which
propagate entirely at very large radii; we intend to provide a more
rigorous justification for its use in another place.

\vskip 0.3 true cm
\noindent
{\underbar {III.2: Behaviour of the Retarded and Advanced Time Functions Near
Infinity}}
\vskip 0.3 true cm
\noindent
The aim of this subsection
is to evaluate the time delay/advance for null geodesics propagating
near infinity.  One begins with a formulation of the conditions of
asymptotic flatness that will be required and which will refer to some
neighbourhood of infinity, ${\cal N} \subseteq {\cal M}$.  It is most
direct to impose conditions on the inverse metric, namely that there
be a quasi-Cartesian coordinate system for $\N$ in which it may be
written as
$$
 g^{\mu\nu}=\eta^{\mu\nu}-4{{m^{\mu\nu}(x^\lambda / r)}
 \over r}+\varphi^{\mu\nu} \ ,
 \eqno{\rm (III.2.1)}
$$
where $r^2=x^2+y^2+z^2$.  The function $m^{\mu\nu}$ describes what
will be called the {\it DC part}
of the asymptotic metric; it is effectively a function on the cylinder
$S^2\times\real$, and will be required to be $C^1$ and bounded.
It is intended that $\varphi^{\mu\nu}$ describe, to
leading order, the radiative part of the metric, so it should fall off
appropriately and satisfy some manner of transversality condition.  It
will be required, specifically,
that $\varphi^{\mu\nu}$ be ${\cal O}(1/r)$,
and that
$\varphi^{\mu\nu} l_{\mu} l_{\nu}$ be ${\cal O}(1/r^{1+\epsilon})$
for some $\epsilon >0$ when $l_\mu$ is taken to be either the out-going,
coordinate-null vector
$l^+_\mu = \partial_\mu (r - t)$ or its in-going counterpart
$l^-_\mu = \partial_\mu (r +t)$.

We now introduce a precise definition for the symbol ${\cal O}(1/r)$. This
definition entails in particular a fall-off which is uniform
on neighbourhoods of compact portions of $\I$.
A function $f$ on $\N$ will be said to be of class ${\cal O}(1/r^\beta)$
with respect to an asymptotic coordinate system
iff, for each function $h(r)$ of the
class $r+o(r)$,
there exists a bound $B>0$ such that
$|f(x)| < B/r^\beta $
for all $x\in\N$ such that $|x^0| < h(r)$.
(By a function of the class $r+o(r)$ is meant one for which
$|h(r)-r| / r \rightarrow 0$ as $r\rightarrow\infty$,
for example the function $h(r) = r + a \log (r/a) + T$.)
Thus, in the cases of $\varphi^{\mu\nu}$ and $m^{\mu\nu}$ one has for
such $x$,
$$
 \eqalignno{\big \vert \varphi^{\mu\nu} \big \vert <& {{\kappa_1}\over r}
 \quad ,&{\rm (III.2.2a)}\cr
 \big \vert \varphi^{\mu\nu}l^\pm_{\mu}l^\pm_{\nu} \big \vert
  <& {{\kappa_2}\over {r^{(1+\epsilon)}}} \quad , &{\rm (III.2.2b)}\cr
\big \vert m^{\mu\nu} \big \vert <&\kappa_3\quad , &{\rm (III.2.2c)}\cr}
$$
for any convenient choice of norm $| \cdot |$ ({\it e.g.} the supremum
over the components in the $x^{\mu}$ basis, the root-sum-square of the
components in this basis, {\it etc.}) and some constants $\kappa_i>0$.

Notice that the conditions for asymptotic flatness are Poincar\'e
invariant: independent of the choice of origin which defines the
radial direction and also independent of the choice of asymptotic rest
frame.  They are weak enough to encompass, for example, the DC
fields and radiation emitted by any astrophysically plausible source
we know of.  (Our particular ``gauge choice'' of asymptotic background
metric has the disadvantage that $r \mp t$ is {\it not} a good
coordinate on $\I^\pm$, but the change of coordinates one would need
to make it so would introduce implicitly defined logarithmic terms
into the asymptotic metric, a complication we wish to avoid.)

The proof to be given in Section IV will proceed entirely within a certain
asymptotic region ${\cal U} \subseteq \N\cup\I$ upon which a single set of
bounds $\kappa_i$ on the fall-off of $h^{\mu\nu}$ holds sway. One may
now estimate the retardation function $\Delta S$ under the assumption
of such fixed bounds.  Since equation ${\rm (III.1.5)}$ is linear in
$h^{\mu\nu}$, the contribution to $\Delta S$ from each of the two
${\cal O}(1/r)$ terms in the metric ${\rm (III.2.1)}$ may be computed
separately.  Hence, the retarded time function at $x$, corresponding
to the generator $\Lambda^+ \subseteq {\cal I}^+$, is
$$
\eqalignno{S^+(x) =&
       t-z+\Delta S^+(x)\quad ,&{\rm (III.2.3a)}
\cr
\Delta S^+(x) =&
   -{1\over 2}\int\limits_t^{\infty}dt'\int d^3x'\ k_{\mu}k_{\nu}
   h^{\mu\nu}(t',{\vec x}')\ \delta^{(3)}({\vec x}-{\vec x}'-(t-t'){\hat
   n})\quad ,
\cr
=& \Delta S^+_{\rm DC}(x)+\Delta S^+_{\varphi}(x)\quad ,&{\rm
(III.2.3b)}
\cr
\Delta S_{\rm DC}^+(x) =&
   2\int\limits_t^{\infty} dt'\int d^3x'\
    {{k_{\mu}k_{\nu}m^{\mu\nu}}\over {r'}}\ \delta^{(3)}
     ({\vec x}-{\vec x}'-(t-t'){\hat n})\quad ,&{\rm (III.2.3c)}
\cr
\Delta S_{\varphi}^+(x) =&
   -{1\over 2}\int\limits_t^{\infty} dt'\int d^3x'\ k_{\mu}k_{\nu}
  \varphi^{\mu\nu}(x')\ \delta^{(3)}({\vec x}-{\vec x}'-(t-t'){\hat n})\quad ,
&{\rm (III.2.3d)}\cr}
$$
with $k^0$ normalised to unity, and the (possibly infinite)
integration constant coming from $C(t_0)$ in (III.1.5) being left implicit in
(III.2.3c).

Anticipating the answer, the DC piece will produce
an unbounded contribution, so it will suffice to show that the
$\varphi^{\mu\nu}$
contribution is bounded, and therefore negligible in the appropriate limit.
It will be useful to characterize null geodesics (in ${\cal N}$, as are all
null geodesics discussed in this subsection) by an {\it impact parameter},
defined in the usual manner with respect to the asymptotic coordinates as
the minimum value of
$$
b^2=x^2+y^2 \quad .\eqno{\rm (III.2.4)}
$$
along the geodesic.

Let
$l_\mu=\pm l_\mu^\pm$, where $\pm={\rm sgn}(z)$, so $l_{\mu}$ corresponds
to the radial null direction, inbound for $z<0$ and outbound for $z>0$, and
with $l_0=-1$.  Define
$$
\eqalign{s_{\mu}=&k_{\mu}-l_{\mu}\quad ,\cr
\Longrightarrow k_{\mu}k_{\nu}\varphi^{\mu\nu}=&l_{\mu}l_{\nu}\varphi^{\mu\nu}
+2l_{\mu}s_{\nu}\varphi^{\mu\nu}+s_{\mu}s_{\nu}
\varphi^{\mu\nu}\quad .\cr}\eqno{\rm (III.2.5)}
$$
Equations ${\rm (III.2.2)}$ may be used to bound the magnitude of each term
in this expression, since the entirety of the path of integration lies within
an asymptotic region where a single set of bounds $\kappa_i$ on the
fall-off of $h^{\mu\nu}$ holds sway.
Hence, using $s_{\mu}=(0,{\vec s})$, one may write that
$$
\eqalign{\vert \Delta S_{\varphi}^+(x) \vert <&
{1\over 2}\int\limits_t^{\infty}dt'
\int d^3x'\ \bigg ( {{c_1(\kappa)}\over {{r'}^{(1+\epsilon)}}} + {{c_2(\kappa)}
\over {r'}}\vert {\vec s}({\vec x}')\vert +{{c_3(\kappa)}\over {r'}}\vert
{\vec s}({\vec x}') \vert^2\bigg ) \cr
&\qquad \delta^{(3)}({\vec x}-{\vec x}'-(t-t'){\hat n})\cr
<&{1\over 2}\int\limits_t^{\infty}dt'
\int d^3x'\ \bigg ( {{c_1(\kappa)}\over {{r'}^{(1+\epsilon)}}} +
{{c_4(\kappa)}\over {r'}}\vert {\vec s}({\vec x}')\vert \bigg )\cr
&\qquad \delta^{(3)}({\vec x}-{\vec x}'-(t-t'){\hat n})\quad .\cr}
\eqno{\rm (III.2.6)}
$$
The $c_i(\kappa)$ are constants determined by the $\kappa_i$ and the choice of
norm. As well, from the definition of $s_{\mu}$, one obtains
$$
\vert {\vec s} \vert^2 =2(1-\vert z\vert /r)\ ,
\eqno{\rm(III.2.7)}
$$
giving
$$
\vert \Delta S_{\varphi}^+(x) \vert < {{c_1(\kappa)}\over
{2b^{\epsilon}}}\int\limits_{{\rm Tan}^{-1}({z\over b})}^{{\pi}\over 2}
({\rm cos}\alpha)^{\epsilon-1} \ d\alpha\ +\ c_5(\kappa) \bigg (
{{\pi}\over 2}-{\rm Tan}^{-1} \big ( {z\over b} \big ) \bigg ) \quad ,
\eqno{\rm (III.2.8)}
$$
with $c_5=c_4/\sqrt{2}$ and with ${\rm tan}\alpha=z'/b$.

A more succinct expression may be obtained by extending the geodesic
back to $t=-\infty$, which yields
$$
\vert \Delta S_{\varphi}^+(x) \vert < {{c_1(\kappa)}\over
{2b^{\epsilon}}}\int\limits_{-{{\pi}\over 2}}^{{\pi}\over 2}
({\rm cos}\alpha)^{\epsilon-1}  \ d\alpha\ +\ c_5(\kappa)\pi\quad .
\eqno{\rm (III.2.9)}
$$
Thus the ``radiative contribution'' to the retardation of a null
geodesic from $x \in \N$ to $\Lambda^+$ is uniformly bounded in $x$,
except possibly in the limit $b\to 0$.  Small $b$ will not be relevant
in the arguments that follow. For now, the chief lesson is that, in
the case of large $b$, the contribution to the phase delay (or
advance) due to $\varphi^{\mu\nu}$ {\it is bounded}, as claimed.

Next, the contribution from the DC piece is required.  Here the
definition
$$
 k_{\mu}k_{\nu} \, m^{\mu\nu}(x^\lambda/r) = m(\tau,\theta,\mu)\ ,
\eqno{\rm (III.2.10)}
$$
will be useful, with $\tau=t/r$, $\mu=z/r$ and ${\rm tan}\theta=y/x$.
The integral ${\rm (III.2.3c)}$ reduces to
$$
\Delta S_{\rm DC}^+(x)=2\int\limits_t^{\infty} dt'\
{{m(\tau',\theta,\mu')}\over {\vert {\vec x}-(t-t'){\hat n}\vert }}
\quad ,\eqno{\rm (III.2.11)}
$$
with the primes in the argument of $m$ denoting evaluation of $\tau$
and $\mu$ at the translated (retarded) value $z-t+t'$ of the
$z$-coordinate ($\theta$ is independent of $z$).  The problem now is
that generally this integral diverges; it is essentially the
$\epsilon=c_4=0$ case of equations (III.2.6--9).  This divergence as
such merely expresses the fact that, with our choice of asymptotic
coordinates, the Minkowskian null lines along which we have evaluated
$\Delta S$ arrive at $\Lambda^+$ at infinitely late (or infinitely
early as the case may be) retarded times --- that is to say, they
actually miss $\Lambda^+$ entirely.  In the well known case of
Schwarzschild, for example, the true out-going null surfaces take the
form $r + a \log(r/a) -t = const.$, which diverges logarithmically
from the null surfaces defined by $\eta_{\mu\nu}$.  As mentioned
earlier, it is because of this effect that (III.1.5) requires a
$t_0$-dependent constant of integration, which reappears as a
``renormalization'' of (III.2.3c) and (III.2.11).

However, what is important here is the {\it difference} in arrival
time at $\Lambda^+$ of null geodesics with different impact
parameters, and (to leading order) such differences can be found
directly from the integral in (III.1.5) or (III.2.11) because the
value of the integration constant drops out. This is the procedure that
will be followed.\footnote{$^{\dag}$}
{Another way to describe this procedure is as follows.  One wants the
difference $S^+(x_0'')-S^+(x_0')$.
To find it, one propagetes null geodesics $\g'$ and $\g''$ from $x_0'$
and $x_0''$ respectively, to $\Lambda^+$ ({\it i.e.} exact null geodesics
with respect to the true metric $g_{ab}$).
When $\g'$ and $\g''$
reach the far asymptotic region, $r\to\infty$,
they are running parallel to $k^\mu$, and the difference in their
$k^\mu$-normalised time of arrival at $\Lambda^+$ is given by the
phase difference $k_\mu (x'-x'')^\mu$, where $x'$ and $x''$ are points
on $\g'$ and $\g''$ in the far asymptotic region.  To leading order
this phase difference is just that due to the difference in starting
phase, $k_\mu (x_0'-x_0'')^\mu$, plus the difference in the
phase delay integrals (III.1.5) computed from $x_0'$ and $x_0''$ to
endpoints of equal $z$ (or $t$) in the far asymptotic region. }
Accordingly, consider a family of null geodesics with the same
initial $z=z_0$ and $t=t_0$, and the same $\theta$, but differing
initial $b$.  One may compute the phase difference between
neighbouring finite-length geodesics specified by $b$ and $b+db$, then
let the final endpoints go to infinity, and lastly integrate the
resulting value of $\partial S^+/\partial b$ to obtain the
$b$-dependence of $S^+$.  (Along all these null geodesics, the same
bounds $\kappa_i$ on the fall-off rates hold.)  Thus, taking two
points along a null line, one with coordinates $(t_0,b,z_0)$, the
other with coordinates $(t,b,z)$ ($t>t_0$ say), and comparing the
phase shift to that along the null line joining $(t_0,b+db,z_0)$ to
$(t,b+db,z)$, one obtains the result
$$
 {{\partial \Delta S^+_{\rm DC}}\over {\partial b}}
   = -{{2zm^+}\over {br}} + {\cal O}(1/b^2)\ .
 \eqno{\rm (III.2.12)}
$$
Here $m^+$ denotes the value of $m$ at the point having coordinates
$(t,b,z)$.  In passing, it may be remarked that the ${\cal O}(1/b^2)$
terms come from two sources, one being the initial endpoint
contribution $2z_0m^-/br_0={\cal O}(1/b^2)$, the other being
$b$-dependence of the arguments of $m(\tau,\theta,\mu)$.

In the limit as the final endpoint goes to infinity, equation ${\rm
(III.2.12)}$ becomes
$$
 {{\partial S^+_{\rm DC}}\over {\partial b}} =
  - {{2M^+}\over b}+ {\cal O}(1/b^2)\quad ,
 \eqno{\rm (III.2.13)}
$$
where $M^+$, the limiting value of $m^+$, is a constant (one might
have expected it to retain the $\theta$-dependence, but $\theta$
parametrises a circle whose angular size at fixed $b$ becomes
negligible as its centre approaches infinity).
In terms of $k^\mu$ one has from (III.2.10) that
$$
  M^+ = \lim m^+ = m^{\mu\nu}(k^\lambda/k^0) \, k_\mu k_\nu \quad .
 \eqno{\rm (III.2.14)}$$

Now take $z_0=0$, in which case (since $M^+$ is independent of
$\theta$) one may treat $S^+_{\rm DC}$ as a function of $b$ alone.
Then,
integrating (III.2.13) and adding in the contribution
$\Delta S^+_\varphi$ (which has already been seen to be bounded, by (III.2.9))
yields
$$
 S^+(b) = -2M^+\log (b/b_0) + {\cal O}(1)\quad ,
 \eqno{\rm (III.2.15)}
$$
where $b_0$ is a constant of integration and ${\cal O}(1)$ indicates
terms bounded in the large $b$ limit.  A similar analysis gives the
advanced time function at $z=0$ (and sufficiently large $b$) as
$$
 S^-(b) = +2M^-\log (b/b_0)+{\cal O}(1)\ ,
 \eqno{\rm (III.2.16)}
$$
with $M^- = m^{\mu\nu}(-k^\lambda/k^0) \, k_\mu k_\nu$.

Finally, for curves that have their initial endpoints on $\Lambda^-$ and
final endpoints on $\Lambda^+$, one may define a {\it time-of-flight}
as the difference between the (retarded) time of arrival and the
(advanced) time of departure.  In terms of this concept one may state
the following lemma, which will be used in the proof in Section IV.

{\narrower\smallskip\noindent
{\underbar {Lemma III.2.1:}}
Let $\gamma$ consist of the union of two null geodesics, one that
joins $\Lambda^-$ to a point $w\in{\cal N}$ with $z=0$ and radial
coordinate $b$, and one that joins $w$ to $\Lambda^+$.  For a fixed
set of bounds $\kappa_i$ in ${\rm(III.2.2)}$, there exist positive
constants $R_0$ and $C_0$ such that, if $\g$ remains always at radii
greater than $R_0$ then, to within a correction of magnitude less than
$C_0$, its time-of-flight from $\Lambda^-$ to $\Lambda^+$ is given by
$$
  -2 (M^+ + M^-)\log \big( b/R_0 \big) \ .
 \eqno{\rm (III.2.17)}
$$
\smallskip}

{\narrower\smallskip\noindent
{\underbar {Proof:}}
Consider a neighbourhood ${\cal N}(\kappa)$ of a compact portion of
infinity on which the inequalities ${\rm (III.2.2)}$ hold with some
triple of {\it fixed} bounds $\kappa=(\kappa_i)$, and let $w\in {\cal
N}$ have $z=0$.  If the radial coordinate at $w$ is $b$, then the functions
$S^{\pm}(w)$ may be computed as in this section, provided the
Minkowskian-null trajectory $z-t=const.$ through $w$ remains within
${\cal N} (\kappa)$, and it is easy to choose ${\cal N}(\kappa)$ so
that this is so.  Then, from ${\rm (III.2.15)}$ and ${\rm (III.2.16)}$,
one has
$$
 S^+(w)-S^-(w)=-2(M^++M^-)\log \big( b/R_0 \big) +{\cal O}(1)\quad .
\eqno{\rm (III.2.18)}
$$
The lemma follows.  $\quad \dal$
\smallskip}

In interpreting the time-of-flight, recall that the normalisations of
$S^{\pm}$ are both fixed by the null vector $k^a$ which determines the
generators $\Lambda^{\pm}$ of ${\cal I}^{\pm}$. It is possible to
define time-of-flight for curves joining arbitrary generators.
Consider for example the generator $\Gamma^+\subseteq{\cal I}^+$
associated to the null vector ${{\partial}\over {\partial t}} +{\rm
cos}\theta {{\partial}\over {\partial z}} +{\rm
sin}\theta{{\partial}\over {\partial x}}$.  The time-of-flight of the
curve that leaves $\Lambda^-$, follows a null geodesic to a point $w$
with coordinates $(b,z)$, then follows another null geodesic to
$\Gamma^+$ is given by
$$
 S^+(w)-S^-(w)=-2(M^++M^-)\log (b/R_0)-2M^+\log \big ( 1+{z\over b}
 {\rm tan}\theta\big )+{\cal O}(1)\quad .\eqno{\rm (III.2.19)}
$$
For non-zero $\theta$, the point $w$ may be adjusted to make this quantity
arbitrarily negative, whence such curves may join arbitrarily late points on
$\Lambda^-$ to arbitrarily early ones on $\Gamma^+$, as remarked in an
earlier footnote.  See also the illustration in fig. (3).

\vskip 0.5 true cm
\noindent
\centreline{\bf IV: The Positivity Proof}
\vskip 0.3 true cm
\noindent
Now it is possible to prove the theorem stated in Section ${\rm II}$.

{\narrower\smallskip\noindent
{\underbar {Proof of Theorem ${\rm II.1.1}$:}}
Choose a future-pointing asymptotic null vector $k^a$, as before.
This vector determines a unique null geodesic generator $\Lambda^+$ of
${\cal I}^+$ and (the past-pointing null vector $-k^a$ determines) a
unique null geodesic generator $\Lambda^-$ of ${\cal I}^-$.  Recall
that, by assumption, there exists a neighbourhood of infinity $\N$
and a quasi-Cartesian
coordinate system $x^\mu$ for $\N$ in terms of which the spacetime
metric $g_{\mu\nu}$ fulfils the asymptotic flatness conditions introduced
in Section III.  Recall also that
$\D \cup \I = [I^-(\I^+) \cap I^+(\I^-)] \cup \I$
is assumed to be globally hyperbolic as a subset of $\widetilde\M$ (fig. (2)).
We seek a ``fastest causal curve'' $\gamma$ from $\L^-$, through $\D$,
to $\L^+$.  The curve $\gamma$ will be constructed as the limit of a
sequence of causal curves $\gamma_n$.

In order to obtain $\gamma_0$, the initial curve of our sequence, let
us begin by choosing $x_0\in\D$ and a causal curve from $x_0$ to
$q_0\in\L^+$.  (Such a curve is easily constructed; one can, for
example, pick $x_0$ in the asymptotic region $\N$ and proceed from
there in a null direction coinciding asymptotically with $k^a$.)
Similarly, we can find some causal curve joining $x_0$ to a point
$p_0\in\L^-$.  All our considerations henceforth will be confined to
the
{\it interval} $\langle\langle p_0,q_0\rangle\rangle
:= J^+(p_0)\cap J^-(q_0)$, by
definition a compact subset of the globally hyperbolic set $\D\cup\I$.
For future reference, let us introduce also a radius $R_0$ large
enough so that all points $x\in\langle\langle p_0,q_0\rangle\rangle$
with (finite) radial
coordinate $r(x)>R_0$ belong to $\N$, and let
${\cal U} = \{ x \in \langle\langle p_0,q_0\rangle\rangle
\, | \, r(x) > R_0 \}$.
(Here we also admit $r(x)=\infty$, so that
$\langle\langle p_0,q_0\rangle\rangle \cap\I \subseteq \L^-\cup\{i^0\}\cup\L^+$
is included in ${\cal U}$.)   Clearly, ${\cal U}$ is open in
$\langle\langle p_0,q_0\rangle\rangle $,
whence its complement, $\K := \langle\langle p_0,q_0\rangle\rangle
\less{\cal U}$,
is a compact subset of ${\cal D}$.

Now what will be called a ``faster causal curve'' than $\gamma_0$ from
$\L^-$ to $\L^+$ will be one which departs (weakly) later {\it and}
leaves (weakly) earlier than $\gamma_0$.  Such a curve will by
definition have its initial and final endpoints in
$\langle\langle p_0,i^0 \rangle\rangle $
and $\langle\langle i^0,q_0\rangle\rangle $
respectively.  In order to home in on a fastest curve, let
us introduce a partial ordering $\le$ on
$\langle\langle p_0,i^0 \rangle\rangle  \times
\langle\langle i^0,q_0\rangle\rangle $ by taking $(p',q') \le (p'',q'')$ iff
$\langle\langle p',q'\rangle\rangle  \subseteq
\langle\langle p'',q'' \rangle\rangle $
({\it i.e.} iff $p' \in J^+(p'')$ and $q'\in J^-(q'')$).  Then, defining $F$
as the set of pairs of points in
$\langle\langle p_0,i^0\rangle\rangle
\times \langle\langle i^0,q_0\rangle\rangle $ which
are joined by a causal curve through ${\cal D}$, let
$(p,q) \in \cl F$
be any element of the closure of $F$ which is {\it minimal} with
respect to $\le$.

By the definition of $p$ and $q$, there exists a sequence of points
$p_0, p_1, p_2,\ldots$ ascending  $\L^-$ to $p$, and a sequence
$q_0, q_1, q_2,\ldots$ descending $\L^+$ to $q$, such that each
pair $(p_n,q_n)$ is joined by a causal curve $\gamma_n$ through
${\cal D}$.  We claim (passing to a subsequence if necessary) that the
curves $\gamma_n$ approach a limit $\gamma$, as depicted in fig. $(4)$.
This follows from the
known fact that, for arbitrary compact subsets $A$ and $B$ of any
globally hyperbolic set, the space ${\cal C}(A,B)$ of causal curves
from $A$ to $B$ is compact.  (Here the topology of ${\cal C}(A,B)$ can
be taken to be the so-called Vietoris topology, as will be shown in a
forthcoming paper$^{(30)}$ wherein the existence of limit curves like
$\gamma$ will be proved by a new method.) Thus, since
$\langle\langle p_0,p\rangle\rangle $ and
$\langle\langle q,q_0\rangle\rangle $ are compact subsets of the
globally hyperbolic set
$\D\cup\I$, the sequence $\{\gamma_n\}$ must possess within ${\cal
C}(\langle\langle p_0,p\rangle\rangle ,\langle\langle q,q_0
\rangle\rangle )$ an accumulation point, $\gamma$, as claimed.  By
construction, $\gamma$ is a fastest causal curve from $p$ to $\L^+$,
in the sense that no causal curve from $p$ (and which enters $\M$) can
arrive earlier on $\L^+$.  In consequence, $\gamma$ must remain always
on the boundary of $J^+(p)$, whence it must be a {\it null geodesic
without conjugate points} (if it enters ${\cal M}$ at
all).$^{(7,8,31)}$

The curve $\g$ will be our self-contradictory null geodesic {\it if}
we can rule out the possibility that the $\g_n$ ``escape to infinity''
(in which case we would have
$\g=\langle\langle p,i^0\rangle\rangle \cup\langle\langle
i^0,q\rangle\rangle  \subseteq \L^- \cup \{i^0\} \cup \L^+$,
which never even enters ${\cal M}$).
In excluding this possibility, one may distinguish two cases:
either an infinite number of the $\g_n$ meet the compact set $\K$
(defined above) or they do not.  In the former case $\g=\lim\g_n$ must
also meet $\K$ since, were it to lie in its complement
$\K' := {\tilde\M}\less\K$ (an open set), the $\g_n$ would eventually
also have to lie in $\K'$ according to the definition of the topology
of ${\cal C}(A,B)$.  Thus $\gamma$ must enter $\M$, as desired.

Passing to the latter case, then, we may assume (by possibly omitting
a finite number of curves from the sequence) that none of the $\g_n$
meet $\K$ or, in other words, that (except for their endpoints on
$\L^\pm$) they lie entirely within the subset $\Omega\subseteq\D$,
$\Omega = \{ x \in \langle\langle p_0,q_0\rangle\rangle
\, | \, R_0 < r(x) < \infty \}$.
But our conditions of asymptotic flatness imply that, within $\Omega$
(which is contained in the compact set
$\langle\langle p_0,q_0\rangle\rangle $),
the inequalities ${\rm (III.2.2)}$ hold  with {\it fixed} bounds
$\kappa_i$, and we can assume that the $R_0$ defining $\K$ has been
chosen at least as large as that occurring in Lemma III.2.1.
The conditions for invoking Lemma ${\rm III.2.1}$ are then met, except for the
fact that the $\g_n$ are not necessarily null geodesics.  To remedy
this, let us find\footnote{$^*$}
{Another approach would be to broaden Lemma ${\rm III.2.1}$ to apply to
arbitrary causal curves.
}
on each $\g_n$ a point $s_n$ at which the $z$-coordinate vanishes, and
having found it, modify $\g_n$ so that it becomes a union of ``fastest
curves'' $\g_n^\pm$ from $s_n$ to $\L^\pm$.  Such fastest curves
$\g_n^\pm$ can be constructed just as before for $\g$ itself, and must
be null geodesics for the same reason.\footnote{$^{**}$}
{Alternatively, their existence follows from the fact that
$\partial J^\pm(s_n)$ must contain a ruling null geodesic through $q_n$
(respectively $p_n$).  }
Since they are fastest, the replacement $\g_n \to \g_n^- \cup \g_n^+$
makes $\g_n$ faster than it was, and accordingly does not change the
fact that the modified endpoints $p_n$ and $q_n$ also converge to the
same limits $p$ and $q$.  Thus we can assume without loss of
generality that each $\g_n$ has the form $\g_n^- \cup \g_n^+$, where
$\g_n^\pm$ is a null geodesic from $s_n$ to $\L^\pm$.  Notice finally
that, again as before, we can also reduce ourselves to the case where
all of the $\g_n$ remain within $\Omega$.

Invoking Lemma ${\rm III.2.1}$, we thus can conclude that,
for $M^+ + M^- < 0$,
the time-of-flight of $\g_n$ from $\L^-$ to $\L^+$ is bounded below by
$$
   -2(M^+ + M^-) \log {b_n\over R_0} - C_0,   \eqno{\rm (IV.1)}
$$
where $b_n$ is the value of $\sqrt {x^2+y^2}$ at $s_n$ and $C_0$ is a
constant.  In other words, the difference between the retarded time
coordinate of $q_n$ and the advanced time coordinate of $p_n$ is at
least that given by expression ${\rm (IV.1)}$.

Assume then that $M^+ + M^- < 0$.  If the $b_n$ diverged to $+\infty$,
then expression ${\rm (IV.1)}$
would eventually become greater than the time-of-flight
between $p_0$ and $q_0$, contradicting the definition of $p_n$ and
$q_n$; hence there is some upper bound $R_1$ which the $b_n$ cannot
exceed.  But this means precisely that the $\g_n$ all meet the compact
set $\K_1$ defined like $\K$, but
with $R_1$ replacing $R_0$ or, to
put it slightly differently, that we can choose $R_0$ large enough so
that all of the $\g_n$, and therefore $\g$ itself, meet the compact
set $\K\subseteq{\cal D}$.

We thus conclude that, unless $M^+ + M^- \ge 0$, $\g$ must enter $\D$,
in which case it (or more precisely its intersection with $\D$) will
be an infinite null geodesic without conjugate points, providing the
desired contradiction.  Appealing to the appendix in order to replace
$M^++M^-$ by $ -2 P\cdot k$, we can conclude, finally, that the total
4-momentum $P^a$ must obey $P_a k^a \le 0$, or, since any future-null
vector could have been chosen as $k^a$, that $P^a$ must be
future-causal ({\it i.e.} future-timelike or -null). $\quad
\dal$\smallskip}

{\narrower\smallskip\noindent
{\underbar {\it Remark:}}
One may wonder what happens when $k\cdot P<0$, such as for a positive
mass Schwarz\-schild metric. In that case, one cannot confine the
$\gamma_n$ to meet any compact subset of spacetime; in fact, the
sequences $\{ p_n \}$ and $\{ q_n \}$ will converge to $i^0$, each
curve $\gamma_n$ having successively larger impact parameter, and
the limit curve $\gamma$ will be the ``null geodesic'' joining $i^0$ to
itself!  (In particular, for any $p\in\I^-$, $I^+(p)$ contains all of
$\I^+$ when $P^a$ is future-timelike.)
\smallskip}

To make a possibly greater appeal to intuition, it may be useful to make
the point another way.$^{(13)}$
The formul\ae$\ $of Section III may be used to evaluate implicitly the
$z$-coordinate at which a given null geodesic with endpoint at
$p\in{\cal I}^-$ meets a particular Cauchy surface parametrised by
$t$, simply by setting $S=constant$.  Then the methods of Section
${\rm III.2}$
yield
$$
z(t)-z_0=t-2\int\limits_{z_0/r_0}^{z/r} d\mu\ {{m(\tau(b,\mu),\theta,\mu)}
\over {1-\mu^2}}\ +{\cal O}(1)\quad ,\eqno{\rm (IV.2)}
$$
where ${\cal O}(1)$ signifies order one (bounded) in $b$.  One now may
ask what would be the shape of the wavefront defined by the meeting of
a surface ruled by null geodesics of constant advanced time with a
Cauchy surface of fixed $t$.  To answer this, compute $(\partial
z/\partial b)_{S^-,t}$. The analysis follows as before. One takes the
limit $t_0\to -\infty$, $z_0\to -\infty$, $\mu_0\to -1$, which places
the vertex of the light cone at $p\in{\cal I}^-$.  To ${\cal O}(1/b)$,
one obtains
$$
 {{\partial z}\over {\partial b}} = 2 {{M^-}\over b} \quad ,
 \eqno{\rm(IV.3)}
$$
which in turn implies that the wavefront formed by the null geodesics
from $p\in{\cal I}^-$ meets the chosen Cauchy surface (say $t=0$) in a
two-surface that is asymptotically of the form
$$
   z=2M^- \log b +{\cal O}(1) \quad .\eqno{\rm (IV.4)}
$$
Similarly, a wavefront of past-directed null geodesics
from $q\in{\cal I}^+$ would also form a logarithmic surface, this time
described by
$$
   z=-2M^+ \log b +{\cal O}(1) \quad ,\eqno{\rm (IV.5)}
$$
obtained by holding the retarded time function constant.
The condition that $q$ be in the causal future of $p$ is the condition
that these two wavefronts meet, for which it suffices that
$M^+ +M^- > 0$ ({\it i.e.} that $k\cdot P <0$).  When this inequality
holds, the wavefronts are ``swept forward near infinity'' and
necessarily overlap there, whence $q$ is actually in the chronological
future of $p$.  When the inequality holds in the opposite direction (so
that $k\cdot P >0$), then the wavefronts are ``swept back near
infinity'' and, if they meet at all, it is not within some
neighbourhood of infinity, which is the key point.  Choosing $p$ and
$q$ so that the wavefronts ``just touch'' gives a ``fastest null
geodesic'' $\g$, as before, and thence the contradiction.
These wavefronts are depicted in fig. (5), with the Minkowski space case given
there as well. Fig. (6) depicts the resulting set ${\cal I}^+ \less I^+(p)$
for $M=0$ (and for $M<0$).

Finally, there is a third version of the argument,
which is somewhat closer
to that of Ref. (12) and which avoids the introduction of causal
curves which are not geodesic.  In this version, one begins with
$p\in\L^-$ and chooses a sequence of {\it finite} points
$p_n\in{\cal D}$ descending a causal curve to $p$.  Because $p_n$ is
not at infinity, the boundary of its future will meet $\L^+$ at $q_n$
(say) and will contain a null geodesic $\g_n$ from $p_n$ to $q_n$ with
no finite pair of conjugate points.  (Proof: One knows that some past
null geodesic ruling $\partial J^+(p_n)$ originates at $q_n$ and can
leave $\partial J^+(p_n)$ only at $p_n$.)  One then deduces, along the
lines of the analysis of Section III, that, for $P\cdot k > 0$, the
initial tangent vectors to the $\g_n$ must avoid a neighbourhood of
the future direction along $\L^-$ in order that the $q_n$ not recede
into the infinite future along $\L^+$ (which they cannot do, since
$J^+(p_n)\subseteq J^+(p_{n+1})$ ).  Obtaining a limiting geodesic
$\g$ as before, one concludes that its tangent is not along
$\Lambda^-$ (whence it enters ${\cal M}$) and that it is also free of
conjugate points (since a limit of conjugate-point-free null geodesics
is also such).  This is the desired contradiction.

\vskip 0.5 true cm
\noindent
\centreline{\bf V: Final Observations}
\vskip 0.3 true cm

\noindent
In the introduction it was mentioned that one potential
application of this theorem would be to semi-classical
gravity.  Another potential application would be to spacetimes
possessing closed timelike curves shrouded within event horizons.
Most singularity theorems break down in the presence of
acausalities, and Finkelstein has suggested that causality violations
might be found instead of singularities, were one to investigate the
interior of certain black holes.  In such a situation there might be
no edgeless spacelike hypersurfaces, but Theorem II.1.1 would still
apply as long as the region outside the hole were globally hyperbolic.

Returning to the semi-classical situation for a moment, there is
a potential {\it extension} of Theorem II.1.1 which would also be of
interest in that context.  Classically, the fact that the energy is
bounded below is often taken to imply the stability of flat space, and
an extension of this fact to semi-classical gravity\footnote{$^*$}
{Once again, the reference here is to stability in the context of the
semi-classical Einstein equations.  It is not only to the more limited
question of the presence or absence of instanton solutions tunneling
to classical states of lower energy, which of course are already ruled
out by the classical positivity of energy.}
would be a step
toward proving stability in the fully quantum case.  However the
result presented here does not deal with the case of zero mass.  For
completeness, one would like such an extension even in the classical
case, of course, but it has a greater potential interest
semi-classically.  There, unless one can rule out the existence of
non-flat solutions of zero energy, one leaves open the interesting
possibility of inequivalent ground states between which tunnelling
would be expected to occur.

It seems that any $P^a=0$ result would be obtained most naturally in
conjunction with a further generalisation, a proof that the 4-momentum
cannot be null.  The reason is that, in both situations, one would be
dealing with the case $P\cdot k=0$ for an appropriate $k^a$; in the $P^a=0$
case any choice of $k^a$ will do, while in the other case one chooses $k^a$
parallel to $P^a$.  For such
a $k^a$, $k^ak^b h_{ab}$ falls off faster than ${\cal O}(1/r)$, say
perhaps as ${\cal O}(1/r^2)$.
If one considers null geodesics originating at $p$ on the generator of
${\cal I}^-$ corresponding to this $k^a$, the result of this fall-off rate
is to modify the relative delay of neighbouring
null geodesics at large $b$. In analogy with equation ${\rm (IV.3)}$, one
obtains
$$
{{\partial z}\over {\partial b}} = {\cal O}(1/b^2) \quad .\eqno{\rm
(V.1)}
$$
When one integrates this, one sees that the relative time delay of
different null geodesics (at least, ones that propagate so as to remain always
near ${\cal I}$) vanishes as ${\cal O}
(1/b)$, so the null geodesics from $p$ now will form a sequence whose
large $b$ limit arrives at ${\cal I}^+$ at finite retarded time $U$ (equal to
the retarded time of arrival one would compute by ignoring $h_{ab}$ and using
the flat metric $\eta_{ab}$).\footnote{$^{**}$}
{A special case of this
behaviour is Minkowski space, where all null geodesics from any fixed $p\in
{\cal I}^-$ arrive at ${\cal I}^+$ at exactly the same retarded time.}
If any of the null geodesics from $p$ with finite $b$ arrive at
earlier retarded times than $U$, then the earliest of these arrivals
is a null geodesic ruling the boundary of $I^+(p)$ and so cannot have
focused, whence the theorem applies. However, it is conceivable that
no geodesic arrives earlier than the limiting time $U$, in which case
there is no apparent contradiction.

This would seem unfortunate, and one might hope that there may be more
subtle contradictions that emerge in this case because it seems
reasonable to hope that the methods applied herein should yield
results known to be true from the
earlier techniques. However, there is also a contrary indication that
perhaps any such extension of the present methods would prove too much!
Specifically, postulate for a
moment the existence of $N$ additional {\it non-compact} spatial dimensions.
Then, roughly speaking, finiteness of the ADM mass requires that the metric
behave as
$$
g_{ab}=\eta_{ab}+{\cal O}(1/r^{1+N})\quad ,\eqno{\rm (V.2)}
$$
which, again in analogy with equation ${\rm (IV.3)}$, yields
$$
{{\partial z}\over {\partial b}} = {\cal O}(1/b^{1+N}) \ ,
\eqno{\rm (V.3)}
$$
even for future-timelike $P^a$.
Again, the integral of this shows that the endpoints of null geodesics of
successively larger $b$ approach a finite limiting value of retarded time on
${\cal I}^+$, precisely mimicking the behaviour in the usual three spatial
dimensions when $P^a=0$.
There is thus a danger that any proof excluding $P^a=0$ in
3+1-dimensions would also erroneously exclude known asymptotically flat
solutions with $P^a \ne 0$ in higher dimensions.

Another possible line of generalisation which might be instructive to
pursue concerns alternative gravity theories, or theories with
compactified extra dimensions.  In some of those theories, positive
energy theorems have been proved, and an attempt to reproduce them
with the methods of this paper would help disclose the extent to which
Theorem ${\rm II.1.1}$ depends on special features of Einstein gravity in
four dimensions.  For example, in 5-dimensional Kaluza-Klein theory,
the total energy is positive for some choices of spatial topology but
of indefinite sign in others, whereas it is not apparent how the
general ingredients entering into the present proof, such as the focusing of
null geodesics, would be affected by the dimensionality, or how they
could be sensitive to the topology in this way.\footnote{$^{\dag}$}
{The zero-energy solution discussed by Witten$^{(32)}$ provides one
interesting example.  Although one might think that a theorem
analogous to ours ought to rule out such metrics (or really the
negative energy solutions with which it is associated$^{(33)}$),
it turns out that spacetime of reference (32) contains a ``bubble of
nothingness which grows at the speed of light'', and our conditions of
asymptotic flatness would not be satisfied in the presence of such a
disturbance.
Kaluza-Klein examples wherein the topological character of ${\cal I}$
can be affected in other interesting ways also exist, such as the
monopole solution, whose twisted character as a circle bundle
would seem to lead to an extended spacetime $\tilde\M$, which is not
even a manifold$^{(34)}$.
}
Similarly, in Kaluza-Klein and other theories, the inertial mass is
typically less closely tied to the ``active gravitational mass'' than
it is in General Relativity, and the purely transverse character of
radiation can also be lost --- both of these being elements on which
the present method of proof directly depends.

Finally, it seems likely this proof may be generalised to
4-dimensional metrics with slower fall-offs rates.  Although the
fall-off conditions assumed herein seem fairly weak, it is hoped that
the method of proof will apply as well to metrics with fall-off rates
as slow as $1/r^{{1\over 2}+\epsilon}$, $0 < \epsilon < 1/2$, on
approach to $i^0$, for such a rate is sufficient to allow for the
definition of a convergent mass integral. With such fall-offs, one
might expect the $b\to\infty$ logarithmic divergences encountered in
Section IV to be replaced by even stronger power law divergences, so
it seems reasonable to hope that the method of proof used would apply
in those cases as well.

\vskip 0.5 true cm
\noindent
\centreline{\bf Acknowledgements}
\vskip 0.3 true cm
\noindent
The authors would like to thank the following institutions:
E.W. thanks NSERC (Canada) for a NATO Science Fellowship during
the earlier phase of the work, the NRC (USA) and R. Hellings of JPL
for a Resident Research Associateship at the JPL
during the later phase of the work,
and Syracuse University Dept. of Physics for hospitality.
R.D.S. and E.W. thank the NSF (USA) for support under grants
PHY 8918388 and PHY 9005690.
Each of the authors have had helpful discussions with several of their
individual colleagues, particularly Abhay Ashtekar, Curt Cutler, John
Friedman, Robert Geroch, Joshua Goldberg, Joseph Samuel, Robert Wald,
and Ulvi Yurtsever, to whom the authors are grateful.

\vskip 0.5 true cm
\noindent
\centreline{\bf Appendix: Superposed Asymptotically Schwarzschild Metrics}
\vskip 0.3 true cm
\noindent

With respect to a flat background metric $\eta_{ab}$, the
Schwarzschild metric may be written in isotropic form as
$$g_{ab}= \biggl [ \bigg ( 1+{M\over {2\rho}} \bigg )^4 - \bigg ( {{1-
{M\over {2\rho}} }\over {1+ {M\over {2\rho}} }} \bigg )^2 \biggl ] U^c
\eta_{ac}U^d\eta_{bd} +\bigg (1+{M\over {2\rho}} \bigg )^4\eta_{ab}\quad ,
\eqno{\rm (A.1)}
$$
where $\eta_{ab}U^aU^b=-1$,
and $\rho^2=\eta_{ab}X^aX^b+(\eta_{ab}X^aU^b)^2$.
The 4-momentum of the solution is $P^a=MU^a$.
By convention, $U^a$ will be future-directed, so a
negative value of $P^0$
means a negative value of $M$.  Asymptotically,
this metric is
$$g_{ab}=\bigg ( 1+{{2M}\over {\rho}} \bigg ) \eta_{ab}
 +{{4M}\over {\rho}} U^c\eta_{ac}U^d\eta_{bd}\quad .
 \eqno{\rm (A.2)}
$$

{}From equation ${\rm (A.2)}$ it is easy to write down the
superposition of an arbitrary number of asymptotic Schwarzschild
metrics indexed by $A$. One obtains
$$
 g_{ab} =  \eta_{ab} +
 2 \sum_A{ M_A {\eta_{ab} + 2 \eta_{ac}U_A^c \eta_{bd}U_A^d }
 \over {\rho_A} } \quad .
 \eqno{\rm (A.3)}
$$
For definiteness, the superposition has been written as a discrete sum,
but more generally one could include a continuous integral as well
(and possibly still more general sorts of superposition).
Because the $M_A$ are not restricted to be positive, a superposition
of this sort may possess any value of $P^a$, including spacelike and
negative timelike ones.

For the sake of deriving equation ${\rm (A.8)}$ below, we assume that a
superposition like ${\rm (A.3)}$ can serve, on approach to spatial
infinity, to describe the most general behaviour of the metric at
${\cal O}(1/r)$ or, in other words, that such a form captures the
asymptotic metric which remains after radiation terms have been
excluded.  Correspondingly, we reason in the main text as if the most
general DC metric can be taken to be of this form, and we limit
the metric perturbation $m^{\mu\nu}/r$ of expression ${\rm (III.2.1)}$
accordingly.  While this could be adopted as part of our definition of
asymptotic flatness, we suspect that such a limitation would be merely
a convenient short-cut which could be avoided at the cost of some
extra work needed to derive (A.8) directly from fall-off conditions
together with the asymptotic field equations.

Now compute $M^\pm$, as defined by equations (III.2.1),
(III.2.10), and (III.2.14), for the metric ${\rm (A.3)}$
and the (future-pointing)
null vector $k^a$.  By definition, this is just the limit as
$\lambda\to\pm\infty$ of ${{-1}\over 4} r h^{ab}(x)\,k_a k_b$,
where $(x-x_0)^a = \lambda k^a$
and $r$ is the radial coordinate of $x$.  Since
$M^+$ is linear in $h^{ab}$, it may be computed for ${\rm (A.2)}$ alone.
Denoting the vector $\partial/\partial t$ by $e^a$ and recalling that
$k \cdot k \ideq \eta_{ab} k^a k^b =0$, one has
$$\eqalign{\rho^2 =& \eta_{ab} x^a x^b + (U\cdot x)^2\cr
        =& (k\cdot U)^2 \lambda^2 + \O(\lambda)\cr
\Longrightarrow \rho =&-\lambda k\cdot U + \O(\lambda^0)\quad ,\cr}
\eqno{\rm (A.4)}$$
and analogously
$$
 r = - \lambda k\cdot e + \O(\lambda^0) \quad .\eqno{\rm (A.5)}
$$
(Here ${\cal O}(\lambda^0)$ means boundedness in $\lambda$.)
Hence, to leading order in $1/r$ and $1/\lambda$, we have
$$
\eqalignno{
 h^{ab}k_a k_b =& (\eta^{ab}+h^{ab})k_a k_b = g^{ab} k_a k_b \cr
 =& g_{ab}k^ak^b
 = \big( 1 + {2M\over \rho} \big) (k\cdot k)
                     + {4M\over \rho} (U\cdot k)^2 \cr
               =& { 4M (U\cdot k)^2 \over -\lambda(U\cdot k)}
               = - 4 {M U\cdot k \over r}  k \cdot e\quad ,&{\rm (A.6)}\cr
 \Longrightarrow M^+ = M^- =&M(U\cdot k)(k\cdot e) = (P\cdot k)(k\cdot e)
 = - k^0 P\cdot k \quad .&{\rm (A.7)}\cr}
$$
Therefore, if $k^0=1$ (as in the main text), then
$$
  M^+ + M^- = -2 P\cdot k \quad ,\eqno{\rm (A.8)}
$$
for ${\rm (A.2)}$, and hence for ${\rm (A.3)}$.
This may be inserted into the results
of Section ${\rm III}$ and used in the proof of Theorem ${\rm II.1.1}$
presented in Section ${\rm IV}$.

In connection with Section III.2, it may be illuminating to
consider an example consisting of the addition of two asymptotic
metrics of the above form. For sake of the example, the result of this
addition will be expressed in a frame where one of the two
constituent metrics is time-independent. In this frame, say that the
other metric is boosted so that it has 4-velocity proportional to $(1,0,
0,V)$ with $\vert V \vert <1$.
Let $R^2=X^kX^k$ with $k\in\{ 1,2,3 \}$. Then the result is
$$\eqalign{ds^2=&\biggl \{ 1+\sum_{A=1,2} {{2M_A}\over R}\bigg [ 1- {{T^2}
\over {R^2}} +{{(T-V_AZ)^2}\over {R^2(1-V_A^2)}} \bigg ]^{-1/2} \biggl \}
\eta_{ab}dX^adX^b\cr
&+{4\over R} \bigg \{ M_1 dT^2
+{{M_2}\over {\sqrt{1-V^2}}}{{(dT^2-2VdZdT+V^2dZ^2)}\over
{\sqrt{(1-T^2/R^2)(1-V^2)+(T/R -VZ/R)^2}}} \bigg \}\quad ,\cr
\quad ,\cr}\eqno{\rm (A.9)}$$
where $V_1=0$ and $V_2=V$.

Consider two cases,
one where the null geodesics of Section ${\rm III}$ propagate perpendicular
to the boost velocity and one where they propagate parallel to the boost
velocity. This corresponds to two different choices of generator of
${\cal I}^-$ on which to place the initial point from which the null geodesics
emanate. One may read off the function $M(T/R,\theta,\mu)$.
The two cases give
$$\eqalignno{
M_{\perp}(T/R,\mu)=&M_1 +{{M_2}\over {\sqrt{1-V^2}}}
\bigg \{ \big ( 1-V^2 \big ) \big ( 1-{{T^2}\over {R^2}}\big ) +\big (
{T\over R}-V {Z\over R} \big )^2 \bigg \}^{-1/2} ,
&{\rm (A.10a)}\cr
M_{\parallel}(T/R,\mu)=&M_1 +{{M_2}\over {\sqrt{1-V^2}}} (1-V)^2
\bigg \{ \big ( 1-V^2 \big ) \bigg ( 1-{{T^2}\over {R^2}}\bigg )\cr
&\qquad +\bigg ( {T\over R}-V {Z\over R} \bigg )^2 \bigg \}^{-1/2} \quad .
&{\rm (A.10b)}\cr}$$
Taking the limits to ${\cal I}^{\pm}$, one obtains
$$\eqalignno{
M^+_{\perp}+M^-_{\perp}=& 2 \bigg ( M_1+{{M_2}\over {\sqrt{1-V^2}}}
\bigg ) \quad ,&{\rm (A.11a)}\cr
M^+_{\parallel}+M^-_{\parallel}=& 2 \bigg ( M_1+{{M_2(1-V)}\over
{\sqrt{1-V^2}}} \bigg ) \quad .&{\rm (A.11b)}\cr}$$

Lastly, since the ADM 4-momentum of this configuration is given by adding the
4-momenta of the two constituent solutions, one boosted and one
stationary, then
$$P^a= \big ( M_1+{{M_2}\over {\sqrt{1-V^2}}},0,0,
{{M_2V}\over {\sqrt{1-V^2}}} \big )\quad ,\eqno{\rm (A.12)}$$
In agreement with (A.8) this results in
$$
\eqalignno{
M^+_{\perp}+M^-_{\perp}=& -2k_{\perp}\cdot P\quad , &{\rm (A.13a)}\cr
M^+_{\parallel}+M^-_{\parallel}=& -2k_{\parallel}\cdot P\quad ,
&{\rm (A.13b)}\cr}
$$
where $k^a_{\perp}$ and $k^a_{\parallel}$ are future-pointing Minkowskian-null
vectors normalised so that $k^0=+1$ in both cases.
The spatial components of $k^a_{\perp}$ are orthogonal to
those of the boost vector $U^a$ while those of $k^a_{\parallel}$ are
parallel.

\vskip 0.5 cm
\noindent
\centreline{\bf References}
\vskip 0.3 true cm
\noindent
\item{(1)}{Schoen, R., and Yau, S.-T., {\it Phys. Rev. Lett.}
{\underbar {43}}, 1457 (1979); {\it Commun. Math. Phys.}
{\underbar {79}}, 47 (1981); 231 (1981).}
\item{(2)}{Witten, E., {\it Commun. Math. Phys.} {\underbar {80}}, 381
(1981).}
\item{(3)}{Nester, J.M., {\it Phys. Lett.} {\underbar {83}}A, 241 (1981);
Parker, T., and Taubes, C.H., {\it Commun. Math. Phys.}
{\underbar {84}}, 223 (1982); Reula, O.,
{\it J. Math. Phys.} {\underbar {23}},810 (1982).}
\item{(4)}{Lee, J., and Sorkin, R.D., {\it Commun. Math. Phys.}
{\underbar {116}}, 353 (1988).}
\item{(5)}{Shapiro, I.I., {\it Phys. Rev. Lett.} {\underbar {13}},
789 (1964).}
\item{(6)}{Will, C.M., {\it Theory and Experiment in Gravitational Physics}
(Cambridge University Press, Cambridge, 1981).}
\item{(7)}{Hawking, S.W., and Ellis, G.F.R., {\it The Large Scale Structure
of Spacetime} (Cambridge University Press, Cambridge, 1973).}
\item{(8)}{Beem, J.K., and Ehrlich, P.E., {\it Global Lorentzian Geometry},
(Marcel Dekker, New York, 1981).}
\item{(9)}{Tipler, F.J., {\it J. Diff. Eq.} {\underbar {30}}, 165
(1978); {\it Phys. Rev.} D{\underbar {17}}, 2521 (1978).}
\item{(10)}{Borde, A., {\it Class. Quantum Gravit.} {\underbar {4}}, 343
(1987).}
\item{(11)}{K\'ann\'ar, J., {\it Class. Quantum Gravit.} {\underbar {8}},
L179 (1991).}
\item{(12)}{Penrose, R., {\it Twistor Newsletter} {\underbar {30}}, 1
(1990).}
\item{(13)}{Sorkin, R.D., and Woolgar, E., {\it Proc. Fourth Can. Conf. on Gen.
Rel. and Rel. Astrophys.}, ed. Kunstatter, G., Vincent, D.E., and Williams,
J.G., p. 206 (World Scientific, Singapore, 1992); Proc. MG6 Meeting held in
Kyoto, Japan, June 1991.}
\item{(14)}{Penrose, R., {\it Essays in General Relativity} (Taub
Festschrift),
p. 1 (Academic Press, New York, 1980).}
\item{(15)}{Geroch, R., reference to be supplied.}
\item{(16)}{Ashtekar, A., and Penrose, R., {\it Twistor Newsletter}
{\underbar {31}} (1990).}
\item{(17)}{Misner, C.W., Thorne, K.S., and Wheeler, J.A., {\it Gravitation}
(W.H. Freeman, New York, 1973).}
\item{(18)}{Wald, R.M., {\it General Relativity} (University of Chicago Press,
Chicago, 1984).}
\item{(19)}{Hawking, S.W., and Penrose, P., {\it Proc. Roy. Soc.
(Lond.)} A{\underbar {314}, 529 (1970).}}
\item{(20)}{Morris, M.S., Thorne, K.S., and Yurtsever, U., {\it Phys. Rev.
Lett.} {\underbar {61}}, 1446 (1988).}
\item{(21)}{Beem, J.K., and Harris, S.G., University of Missouri-Columbia
and St. Louis University preprint (1992); Proc. MG6 Meeting held in Kyoto,
Japan, June 1991.}
\item{(22)}{Hawking, S.W., {\it Phys. Rev} D{\underbar {46}}, 603 (1992).}
\item{(23)}{Roman, T.A., {\it Phys. Rev.} D{\underbar {33}}, 3526
(1986); D{\underbar {37}}, 546 (1988).}
\item{(24)}{Wald, R., and Yurtsever, U., {\it Phys. Rev.} D{\underbar {44}},
403 (1991); Yurtsever, U., {\it Class. Quantum Gravit.} {\underbar {7}},
L251 (1990).}
\item{(25)}{Klinkhammer, G., {\it Phys. Rev.} D{\underbar {43}},
2542 (1991).}
\item{(26)}{Ford, L.H., and Roman, T.A., Tufts Institute of Cosmology
preprint (1992);
Ford, L.H., {\it Phys. Rev.} D{\underbar {43}}, 3972 (1991); Ford, L.H., and
Roman, T.A., {\it Phys. Rev.} D{\underbar {41}}, 3662 (1990);
Ford, L.H., {\it Proc. Roy. Soc. Lond.} A{\underbar {364}}, 227 (1978).}
\item{(27)}{Penrose, R., {\it Perspectives in Geometry and Relativity}
(Hlavaty Festschrift), ed. B. Hoffmann, p. 259 (Indiana University Press,
Bloomington, 1966).}
\item{(28)}{Pleba\'nski, J., {\it Phys. Rev.} {\underbar {118}}, 1396 (1960).}
\item{(29)}{Rindler, W., and Penrose, P., {\it Spinors and Spacetime}, Vol. II
(Cambridge University Press, Cambridge, 1986).}
\item{(30)}{Sorkin, R.D., and Woolgar, E., preprint (1992).}
\item{(31)}{For an interesting proof of the null geodesicity of $\gamma$,
but not the absence of conjugate points along $\gamma$, see Nityananda, R.,
and Samuel, J., {\it Phys. Rev.} D{\underbar {45}}, 3862 (1992).}
\item{(32)}{Witten, E., {\it Nuc. Phys.} B{\underbar {195}}, 481 (1982).}
\item{(33)}{Brill, D., and Pfister, H., {\it Phys. Lett.} B{\underbar {228}},
359 (1989).}
\item{(34)}{Sorkin, R.D., {\it Phys. Rev. Lett.} {\underbar{51}}, 87 (1983).}
\par\vfill\eject
\noindent
\centreline{\bf Figure Captions}
\vskip 0.3 true cm
\noindent
We regret we cannot provide electronic versions of the figures.
\item{Fig. (1a):}
{Focusing of null geodesics.  The geodesics enter the crossing
region and leave the boundary of the future of the initial point,
{\it en route} to the conjugate points they will
eventually encounter when they reach the caustics.}
\item{Fig. (1b):}
{Focusing of null geodesics. A single pair of geodesics is
shown. At $q$, they cross and leave the boundary of the future of $p$, but
they must obey local causality, so they remain (for awhile) on the boundary of
the future of $q$. They will eventually reach the caustics.}
\item{Fig. (2):}
{An example of the Domain of Outer Communications of a
particular asymptotic region of some solution.}
\item{Fig. (3):}
{A conformal compactification with $i^0$ properly displayed as
a single point.  Spacetime is exterior to the cones which represent infinity.
{}From $p\in \Lambda^-$, one may reach any point of any generator of
${\cal I}^+$,
{\it except} $\Lambda^+$, by a causal curve in spacetime produced by
smoothing out the curve that is the union of $\Lambda^-$ with $\{ i^0 \}$ and
with the appropriate generator of ${\cal I}^+$. Such a smooth causal curve in
spacetime is displayed as a dotted line here joining $p$ to $q'$.}
\item{Fig. (4):}
{The sequence of curves $\{ \gamma_n \}$ used in the proof of
Theorem ${\rm II.1.1}$, for the case $M\le 0$.}
\item{Fig. (5a):}
{The $t=0$ surface in Minkowski space. In flat space, parallel lines,
here defined by null geodesics incident on plane wavefronts, meet at
infinity and so define the points of ${\cal I}^{\pm}$. If the two
2-plane wavefronts shown lying in this Cauchy surface are truly
parallel, it is clear they define points of ${\cal I}^{\pm}$ that are
not in causal contact with each other. If they are not parallel, they
must meet somewhere, and causal contact is established. This gives
rise to the ``missing-half-generator'' of fig. (6).}
\item{Fig. (5b):}
{A spacelike surface when $M>0$.
Now curvature bends geodesics en route to/from
${\cal I}$, producing ``logarithmic wavefronts'' that always meet. The
augmentations that appear on the wavefronts are the result of the focusing
depicted in fig. (1).}
\item{Fig. (5c):}
{A spacelike surface when $M<0$. The logarithms of fig. (5b)
are reversed. Again, wavefronts may fail to meet. Because the fall-off near
infinity is only logarithmic in the impact parameter, only one half-generator
of ${\cal I}^+$ is excluded from the future of any chosen point in
${\cal I}^-$,
as depicted in fig. (6).}
\item{Fig. (6):}
{Compactified Minkowski space displayed conventionally,
so that $i^0$ is not
correctly represented as a single point. This diagram is also appropriate
to $M<0$ space.
The accented half-generator of ${\cal I}^+$ is the set of all
points in ${\cal I}^+ \less I^+(p)$. Note that $i^0$ is incorrectly
represented --- it should be a single point.}
\par\vfil\eject\end